# Artificial Pancreas Implantables – How Healthcare Professionals May Deal With DIY Bio Cases

Austin James[1], Xavier-Lewis Palmer[2], Lucas Potter[2], Celisha Oscar[3]
[1]*University of East London, London, UK*
*Email: hello@theaustinjames.com*
[2]*BIOSView Labs, Ohio, USA*
[3]*Howard University, DC, USA*

*Abstract*—**Automated insulin delivery (AID) and artificial pancreas systems increasingly serve as safety-critical cyber-physical technologies in clinical care, integrating sensors, algorithms, software, and insulin-delivery hardware to automate a life-sustaining therapy. While regulated commercial systems are supported by formal approval pathways, manufacturer governance, and post-market surveillance, clinicians are also encountering patients who rely on do-it-yourself (DIY) artificial pancreas systems that operate outside conventional regulatory and institutional control structures. This paper examines how routine clinical handling practices intersect with cyberbiosecurity risk across both regulated and DIY AID systems. When insulin delivery systems are fundamentally reconfigured into a bespoke AID system, with the patient-user becoming the primary threat vector by assuming manufacturer-level roles without mandated governance, the entire ecosystem of stakeholders is placed in legal and clinical uncertainty.**

## 1. Methods and Contribution

We examine clinical handling themes such as eligibility for use, override authority, alarm management, data validation, and transitions of care from a socio-technical systems perspective. These aspects relate to attacks on integrity, availability, and authorisation relevant to patient safety. We highlight that cyberbiosecurity risks in artificial pancreas systems stem from mismatches between clinical responsibility and technical authority, especially in DIY contexts with diffuse software provenance. Examining national position statements from Australia, Canada, and the UK reveals varied institutional responses to managing risk without endorsing unregulated technology. We suggest that regulatory frameworks can unintentionally encourage DIY adoption, creating further mismatches in responsibility and authority. Our proposed minimal clinical cyber-safety handling bundle focuses on harm containment, clear governance, and defensible decision-making instead of optimising devices. Finally, we identify critical gaps in research regarding failure modes, incident reporting, and DIY system governance, emphasising the importance of grounding cyberbiosecurity analysis in everyday clinical workflows to frame it as a governance challenge for patient safety. What follows is a descriptive position paper that sheds light on this niche with the aim of fostering further discussions.

## 2. Introduction and Background

Automated insulin delivery (AID) and artificial pancreas systems are a class of safety-critical cyber-physical medical technologies that integrate continuous glucose sensing, control algorithms, software interfaces, and insulin-delivery hardware to modulate a life-sustaining therapy in real time. Over the past decade, these systems have transitioned from experimental research platforms to widely deployed clinical tools, with growing evidence supporting their effectiveness in improving glycaemic outcomes in people with type 1 diabetes [1], [2]. As adoption has increased, clinicians have been required to manage the operational behaviour of complex, software-driven systems embedded within routine care workflows. This complexity is potentially further amplified by the emergence of patient-built or open-source ("DIY") artificial pancreas systems, which are increasingly reported in real-world clinical contexts despite operating outside conventional regulatory approval pathways. In such cases, healthcare professionals may retain clinical responsibility for patient outcomes while lacking formal authority over device design, software provenance, update mechanisms, or security controls. This misalignment between clinical duty and technical governance introduces distinct safety and cyberbiosecurity challenges, situating artificial pancreas systems at the intersection of medical device regulation, professional policy-making, cybersecurity risk, and the practical tensions of patient-led innovation in chronic disease management [3]. Section 3 provides a detailed description of the topology of these systems and their clinical implications.

## 3. Typologies of Artificial Pancreas Systems and Their Clinical Implications

### 3.1. Regulated Artificial Pancreas and AID Systems

Regulated AID systems are developed, validated, and deployed within formal medical device regulatory frameworks, which define requirements for safety, effectiveness, post-market surveillance, and manufacturer accountability. Commercially available systems combine approved continuous glucose monitors, insulin pumps, and proprietary control algorithms that have undergone clinical evaluation through randomised trials and real-world performance studies. These systems are generally classified as hybrid closed-loop technologies because user input is still required for activities such as meal announcements and system initialisation. At the same time, insulin delivery is continuously modulated in the background [4]. From a clinical perspective, regulatory approval provides clinicians with standardised documentation, manufacturer-supported training materials, defined update pathways, and clear liability boundaries, all of which shape institutional policies governing inpatient and outpatient use. Despite these formal safeguards, regulated AID systems still introduce nontrivial handling complexity in clinical environments. Tanenbaum et al (2025). discuss the behavioural, educational, and practical challenges associated with onboarding and maintaining AID systems [5]. In addition, Phillip et al. (2023) note that the safe and effective use of AID systems depends on adherence to manufacturer-specified behaviours [6]. Actions such as fictitious carbohydrate entries, post-meal bolusing, or overriding algorithmic recommendations can destabilise glucose control and increase hypoglycaemia risk. Studies and consensus statements underscore challenges pertaining to clinician familiarity, alarm burden, integration with hospital workflows, and decision-making during acute illness or procedures [7], [8], [9], [10], [11]. Furthermore, regulatory approval does not eliminate residual risk associated with software-driven automation, particularly when systems are used outside the conditions under which they were validated or when clinical context diverges from the assumptions underlying ambulatory use. As a result, professional guidance increasingly frames regulated AID systems as safety-critical systems requiring active clinical governance, role clarity, and contingency planning instead of set-and-forget" technologies [2]. These characteristics form the baseline against which the clinical and cyberbiosecurity implications of non-regulated and DIY systems can be meaningfully contrasted.

### 3.2. DIY and Non-Regulated Artificial Pancreas Systems

In parallel with the development of regulated AID technologies, patient-led innovation has produced a range of DIY and open-source artificial pancreas systems that operate outside formal regulatory approval pathways. These systems typically integrate commercially available continuous glucose monitors and insulin pumps with community-developed control algorithms, smartphone applications, and cloud-based services, relying on informal peer support networks rather than manufacturer-backed training or maintenance structures. Empirical studies documenting real-world use indicate that DIY systems are adopted by a motivated subset of individuals with type 1 diabetes, often driven by unmet clinical needs, perceived performance advantages, or a desire for greater autonomy over system behaviour [15], [16]. Reported glycaemic outcomes in observational cohorts are often comparable to those achieved with regulated systems, but these findings are derived from self-selected populations and lack the procedural safeguards of controlled trials. From a clinical perspective, DIY artificial pancreas systems introduce a distinct governance challenge. While clinicians may encounter patients whose glycaemic stability is demonstrably influenced by DIY automation, they typically lack access to authoritative documentation, validated update pathways, or clear mechanisms for auditing software integrity and algorithmic logic. Qualitative research highlights clinician uncertainty about legal responsibility, ethical obligations, and the extent to which engagement with DIY systems constitutes endorsement rather than harm reduction [17]. Unlike regulated devices, DIY systems break down the usual lines between patient, developer, and operator. This gives the patient more technical authority, but the clinical responsibility for outcomes stays mostly the same. This divergence is critical for understanding why cyberbiosecurity considerations such as provenance, trust, and configuration control become more salient in DIY contexts, even in the absence of malicious threat models. This conflation in responsibility, with the patient as user, integrator, security administrator, and data curator, creates what we term the 'user-as-accidental-threat' paradox. Unlike regulated systems, in which security governance is handled by manufacturers, DIY systems concentrate all security functions within the patient-user. This makes them the single point of security failure through structural inevitability: well-intentioned actions (misconfigured updates, algorithm tweaks, and delayed patches) become primary threat vectors to system integrity and availability. A recent development complicates this divide. In 2023, Tidepool Loop became "the first open-source automated insulin delivery. . . mobile application. . . recognised as safe and effective by a regulatory body" [18]. This illustrates that the governance challenges seen in DIY AID arise less from open-source code itself than from the absence of an accountable entity capable of maintaining documentation, ensuring version control, and providing clinical evidence.

### 3.3. Clinical Handling Themes Across System Types

Across both regulated and DIY artificial pancreas systems, clinicians encounter recurring handling themes that shape day-to-day safety management. Assumptions differ markedly across system types. A primary consideration is determining eligibility for continued use of a personal AID system during clinical care, which typically depends on patient capacity for self-management, clinical stability, and the ability to ensure reliable insulin delivery. Professional guidance for regulated systems emphasises structured assessment and predefined criteria for continuation or conversion to hospital-managed insulin therapy, particularly in the context of acute illness like diabetic ketoacidosis [19]. For DIY systems, similar clinical judgements must be made in the absence of standardised documentation or institutional endorsement, increasing reliance on clinician discretion and patient mediation. Override authority and decision-rights allocation constitute a second critical theme. Safe use of AID systems requires clarity regarding who may adjust settings, deliver boluses, or suspend automation, especially in inpatient environments where multiple actors interact with the technology. There is value in investigating if and how hospital pumps and AID use highlight that ambiguity in control roles is a recognised source of error and liability concern [20]. Alarm handling and data validation further complicate clinical workflows, as clinicians must distinguish sensor artefact from true dysglycaemia and device malfunction, often amid alarm fatigue and competing clinical priorities. Consensus statements on continuous glucose monitoring in hospital settings emphasise the need for confirmatory testing and defined escalation pathways when automated data conflict with clinical context [21]. Finally, transitions of care, including procedures, imaging, and perioperative periods, expose vulnerabilities related to insulin interruption, unsafe restarts, and loss of situational awareness, underscoring the need for contingency planning irrespective of whether the system is regulated or DIY. Together, these themes illustrate that the clinical handling of artificial pancreas systems is less governed by algorithmic sophistication than by the robustness of human–system coordination across varying governance regimes.

TABLE I. GOVERNANCE AND ACCOUNTABILITY DIFFERENCES BETWEEN REGULATED AND DIY AID SYSTEMS

| Dimension | Regulated AID Systems (Commercial Systems and Tidepool Loop) | DIY AID Systems (Patient-Managed) |
|---|---|---|
| Regulatory Status | Subject to MDR (EU), UK MDR, and FDA clearance; classified as Class II/III depending on jurisdiction | No regulatory classification; operate entirely outside medical device frameworks |
| Accountable Entity | Manufacturer holds full lifecycle responsibility (design, risk management, updates, PMS) | No accountable manufacturer; responsibilities are not reassigned and effectively vanish |
| Security Governance | Secure-by-design, SBOM, validated update pathways, documented risk file [12], [13] | Informal update channels, unverifiable software provenance, no SBOM or formal risk management |
| Transparency Model | Controlled transparency (e.g., Tidepool: open-source algorithm but obfuscated partner-device interfaces) | Full transparency of code, but no validated release branch, no controlled distribution |
| Clinical Requirements | Must demonstrate safety, effectiveness, and substantial equivalence | No clinical trials, no formal evidence generation, no incident reporting |
| Post-Market Surveillance | Mandatory PMS, incident reporting, version control, and recall mechanisms [12] [13] | No PMS, no reporting pathways, no structured monitoring |
| Clinician Trust Model | Institutional trust anchored in regulatory assurance and manufacturer accountability | "Responsibility without control": clinicians must support therapy without authoritative system knowledge [14] |
| Risk Profile | Managed through regulatory controls and manufacturer governance | Shifted to patients and clinicians, integrity, availability, and safety risks are unbounded |
| Example | Control-IQ, CamAPS FX, Tidepool Loop (FDA-cleared) | Loop, OpenAPS, AndroidAPS |

# 4. Regulatory and Policy Environment

## 4.1. Multijurisdictional Medical Device Frameworks

**4.1.1. The Manufacturer-Centric Governance Model.** Modern regulatory frameworks for medical devices – including the European Union Medical Device Regulation [12], UK MDR, and U.S. FDA cybersecurity guidance – establish a clear chain of accountability that begins and ends with the manufacturer. This model effectively removes security management from clinical environments, creating a bounded, if imperfect, basis for institutional trust in commercial AID systems. Tidepool Loop demonstrates that open-source AIDs can also be brought under this manufacturer-centric model. Its FDA clearance was "an important milestone. . . based on an algorithm initially designed and developed by users" [18]. Tidepool achieved this by assuming full lifecycle governance and demonstrating substantial equivalence to Control-IQ. Its implementation remains "largely open-source, but the components allowing access to partner devices are likely to be obfuscated" [22], illustrating a hybrid model where transparency coexists with tightly controlled safety-critical elements.

**4.1.2. The Governance Vacuum in Patient-Integrated Systems.** When patients construct or modify DIY AID systems, they operate outside this manufacturer-centric model. Crucially, the security governance obligations defined by regulations – maintaining risk management files, ensuring update integrity, and reporting incidents risk abandonment Empirical studies of DIY users consistently identify motivations, including a desire for greater autonomy, customisation, and perceived performance advantages over available commercial systems [15], [17]. This suggests that while regulatory frameworks ensure baseline safety, they may not fully accommodate the range of patient needs and preferences that drive DIY adoption.What results is a structural misalignment: clinicians assume responsibility for outcomes, while the regulatory mechanisms designed to ensure device safety are absent. Tidepool Loop's case can suggest that the governance challenges in DIY AIDS arise from the absence of an accountable entity to maintain documentation, ensure version control, and provide clinical evidence.

**4.1.3. From Regulatory Gap to Clinical Challenge.** This governance vacuum transforms regulatory non-compliance into a direct clinical risk management problem. Healthcare institutions serve patients whose glycaemic stability may depend on systems with unverifiable software provenance, informal update mechanisms, and no formal incident-reporting pathways. The clinical consequence is that clinicians must manage safety-critical therapy without authoritative access to the systems governing it [14]. In the EU, automated closed-loop systems are classified as Class III, triggering MDR requirements—full technical documentation, device-specific clinical evidence, and equivalence contracts—that decentralised open-source groups generally lack the resources or legal structure to satisfy [18]. To clarify the structural divergence between regulated and DIY AID systems, Table I summarises governance, accountability, and clinical implications across key regulatory dimensions.

### 4.2. National Policy Responses to the DIY Phenomenon

The emergence of DIY AID systems has prompted formal responses from national healthcare bodies, creating a distinct policy spectrum for navigating this regulatory anomaly. The Australian Diabetes Society advocates for pragmatic containment, establishing clear ”red lines” to separate clinical care from technical support [23]. Diabetes Canada emphasises cautious education, steering patients toward approved alternatives while acknowledging autonomy [24]. UK approaches indicate an evolution toward managed engagement. Breakthrough T1D UK (formerly JDRF) acknowledges the significant glucose improvements reported by users while maintaining that clinical support must be provided, without the clinician assuming responsibility for the DIY technology itself [25]. Furthermore, a consensus statement in The British Journal of Diabetes notes that although systems are not “approved”, clinicians have an ethical duty to support patients to ensure safety and avoid abandonment [26]. This is supported by international expert consensus—including that of Diabetes UK—which provides a framework for healthcare professionals to ethically engage with “open-source” users amid the existing clinical-legal paradox [27]. These policies transform clinicians into regulatory boundary enforcers, creating compartmentalised trust that acknowledges patient agency while maintaining professional safeguards. These divergent approaches can be summarised as a spectrum of responses, with each implementing distinct compensatory controls to address the regulatory vacuum. This divergence is seen in Table II below: This comparison illustrates how national responses navigate the “responsibility without control” dilemma, with clinicians playing a pivotal role in ensuring that boundaries are respected.

### 4.3. Author Responses to the DIY Phenomenon

DIY AID systems reflect structural pressures—cost, inequitable access, and slow commercial innovation—that current regulatory frameworks do not address. We propose a governance model distinguishing safety-critical functions (requiring regulatory oversight) from customizable components (remaining open-source) and recommend that manufacturers discontinuing support release safety-critical code under an open-source license to prevent device abandonment.

## 5. Clinical Workflows as Cyber-Physical Attack Surfaces

### 5.1. Clinical Workflow as a Safety-Critical Socio-Technical System

As established in Section 3.3, AID systems function within safety-critical socio-technical workflows where eligibility determinations, override authority, alarm interpretation, and care transitions create recurring coordination challenges. Viewed through a socio-technical lens, these handling themes reflect the interdependence of technical integrity, human judgment, and organisational control [28]. Automation redistributes cognitive work and obscures system states, creating dependencies on shared situational awareness and clear role delineation, conditions where small deviations can cascade into clinically significant harm [29]. Recognising clinical workflow as a safety-critical socio-technical system therefore provides a necessary foundation for analysing cyberbiosecurity risks, as it foregrounds the interdependence of technical integrity, human judgment, and organisational control in determining the real-world safety of artificial pancreas technologies whose physiological effects are mediated through software-driven control and data integrity.

### 5.2. Provenance, Trust, and Control Boundaries in Artificial Pancreas Systems

The security of artificial pancreas systems depends on software provenance, trust, and control boundaries, as these dimensions determine who can legitimately modify system behaviour and who bears responsibility when failures occur. In regulated AID systems, provenance is anchored in manufacturer-controlled software development lifecycles, validated update pathways, and post-market surveillance obligations, providing clinicians and institutions with a bounded, if imperfect, basis for trust [30]. By contrast, DIY and non-regulated systems rely on community-developed code, informal validation practices, and user-managed updates, dissolving traditional distinctions between developers, operators, and patients. Empirical studies of DIY AID use highlight that clinicians often lack visibility into algorithmic logic, software versioning, and configuration changes, yet remain responsible for managing clinical consequences when automation influences glycaemic outcomes [17]. From a socio-technical perspective, this redistribution of control creates asymmetric trust relationships in which clinicians must make safety-critical decisions without having authoritative access to the system’s internals. Research on trust in automated and cyber-physical systems demonstrates that safety degrades when operators must take responsibility without control [14]. In artificial pancreas systems, this condition is most acute in DIY contexts, where software provenance cannot be independently verified, and configuration changes may occur outside clinical awareness. These trust and control boundary mismatches transform routine clinical tasks, such as deciding whether to continue automation, suspend insulin delivery, or revert to manual therapy, into implicit risk management decisions with cyberbiosecurity implications. Explicitly incorporating provenance and control boundaries into the analysis clarifies that cyberbiosecurity concerns in artificial pancreas systems encompass external malicious threats and structural vulnerabilities arising from fragmented governance and opaque system evolution.

TABLE II. *Comparative Summary of National Policy Responses to DIY AID Systems*

| Jurisdiction | Core Stance | Key Mechanism | Implied Trust Model | Primary Compensatory Control |
|---|---|---|---|---|
| Australia (Diabetes Australia) | Pragmatic Containment | Establishes clear "red lines" separating clinical care from technical support. | Compartmentalised but Distant | Mitigates availability and integrity risks by refusing engagement with unknown system states, forcing fallback to hospital protocol. |
| Canada (Diabetes Canada) | Cautious Engagement and Steerage | Acknowledges patient autonomy while steering users toward approved, regulated alternatives. | Guided | Manages authorisation and awareness risks by educating within a framework that prioritises transition to regulated systems. |
| United Kingdom (Breakthrough T1D UK, Expert Consensus) | Managed Engagement | Supports patient safety without assuming responsibility for DIY technology; provides ethical frameworks for professional engagement. | Compartmentalised but Engaged | Addresses integrity and safety risks through structured clinical support that decouples patient care from device endorsement. |

### 5.3. Mapping Clinical Functions to CyberBiosecurity and Safety Attack Surfaces

Mapping clinical handling functions to cyberbiosecurity and safety attack surfaces enables a structured analysis of how routine care practices intersect with vulnerabilities in artificial pancreas systems. Drawing from Section 5.1, from a systems-safety perspective, clinical decisions, such as eligibility to continue automation, assignment of override authority, and management of alarms, serve as control points that mediate risks arising from integrity, availability, and authorisation failures in cyber-physical systems. In regulated AID systems, the range of plausible system states and changes is partially constrained by regulated software lifecycle and maintenance practices (often operationalised through IEC 62304-aligned processes and quality controls), which provide clinicians and institutions with a bounded, though not complete, basis for trust in software provenance and change management [31]. Nevertheless, cybersecurity scholarship demonstrates that implantable and body-adjacent medical technologies, including insulin delivery ecosystems, can remain vulnerable to failures in authentication, communication security, and privacy protections, which may directly lead to patient harm [3]. In DIY and non-regulated systems, these same clinical functions map onto broader and less bounded attack surfaces because software provenance, update governance, and configuration control are often diffuse, user-managed, or community-mediated. This shifts safety assurance toward compensatory clinical and organisational controls, such as stricter continuation criteria, conservative validation practices, and an earlier fallback to hospital-managed therapy rather than reliance on manufacturer-validated lifecycle governance. Foundational analyses of security and safety in implantable medical devices underscore that when update pathways, auditability, and responsibility structures are fragmented, clinical trust necessarily becomes conditional and "imperfect", with heightened reliance on transparency and post hoc reconstruction after adverse events [30]. Framed in this way, mapping demonstrates that cyberbiosecurity risks in artificial pancreas systems also arise from structural mismatches between clinical responsibility and technical control, particularly in DIY contexts. The mapping reveals how policy responses serve as compensatory controls to fill gaps in the regulatory infrastructure. For example, Australia's restriction on any form of technical support mitigates availability risks arising from unknown software states, while awareness and mandatory point-of-care verification help prevent unintended changes that might jeopardise the system's integrity.

## 6. Navigating the Governance Challenge

### 6.1. Implications for Clinical Governance and Duty of Care

Clinical governance for artificial pancreas and AID systems depends on aligning clinical responsibility with decision-making authority, documentation, and institutional oversight. In settings where patients use their own diabetes technologies, established governance practices emphasise clear eligibility criteria for continuation, explicit allocation of decision rights (including who may adjust settings or suspend automation), and predefined fallback pathways to hospital-managed insulin therapy when safety cannot be assured [32]. These principles are designed to manage risk in regulated systems, where device behaviour, software updates, and accountability structures are at least partially legible for clinicians and institutions. However, their application becomes substantially more complex in the context of DIY artificial pancreas systems, where clinicians may have limited visibility into software provenance, algorithmic logic, or configuration changes while remaining professionally responsible for patient outcomes. Clinical ethics may typically analyse this situation as a dilemma of responsibility without control, in which clinicians must choose between disengagement, which may expose patients to avoidable harm, and engagement, which risks being construed as an endorsement of an unapproved system [33]; it is worth acknowledging that the world of ethics relating to this subject transcends the clinical themes covered in this material and merits a paper onto itself for a judicious treatment. Legal scholarship further underscores the ambiguity surrounding clinician liability in such cases, noting that existing regulatory frameworks offer little guidance on whether advising, documenting, or accommodating DIY system use constitutes "prescribing" or authorising an unapproved medical device [34]. Empirical studies of healthcare professionals' perspectives suggest that, in practice, clinicians respond to this uncertainty by adopting more conservative governance strategies, including heightened documentation, restricted adjustments to device settings, and earlier recourse to institutional insulin protocols when clinical conditions change [35]. Collectively, these findings indicate that governance for DIY artificial pancreas use is less about formal control of technology than about risk containment through role clarity, transparency, and defensible clinical decision-making under conditions of constrained technical authority.

TABLE III. *Mapping Clinical Workflow Security From Risk to Response*

| Phase | Clinical Task | DIY Risk | Attack Surface | Policy Response |
|---|---|---|---|---|
| Admission | Eligibility Assessment | Unknown Software State | Availability | Policy Response: "Red line" - no technical support (AU) |
| Monitoring | Data Validation | Algorithmic Data Curation | Integrity | Mandatory point-of-care verification |
| Care Transition | Override Authority | Ambiguous control during handoff | Authorisation | Policy Response: Documented decision rights |
| Discharge | Update Management | Unpatched Vulnerabilities | Availability and Integrity | Policy Response: Awareness without endorsement (CA/UK) |

### 6.2. Toward a Minimal Clinical Cyber-Safety Handling Bundle

A minimal clinical cyber-safety handling bundle focuses on harm containment and defensible governance, rather than optimizing artificial pancreatic systems. This includes establishing clear criteria for the continuation or discontinuation of personal AID systems, along with predefined fallback pathways to hospital-managed insulin therapy during clinical instability or loss of patient capacity. This approach is consistent with inpatient guidance on the use of personal diabetes technologies, which frames continuation as conditional rather than presumptive [32]. Second, decision rights regarding insulin delivery, parameter modification, and suspension of automation should be clearly assigned and documented at the point of care, treating control over automated dosing as a safety-critical function analogous to medication ordering. A third component concerns data integrity and alarm governance. Given the reliance of AID systems on continuous telemetry, clinicians require defined protocols for alarm triage and data validation, including circumstances in which confirmatory point-of-care testing is required and thresholds for reverting to manual control when data reliability is uncertain. Consensus guidance on hospital use of continuous glucose monitoring underscores that such practices are essential to mitigating automation-related risk, independent of device provenance [21]. Finally, baseline cyber-hygiene measures such as limiting undocumented configuration changes during admission, maintaining awareness of controller access points, and preserving relevant device state information for auditability should be integrated into routine clinical documentation rather than treated as separate technical tasks. Importantly, this bundle is not intended to function as a comprehensive cybersecurity framework or as an endorsement of unregulated systems; rather, it represents a pragmatic translation of established clinical safety practices into a form that acknowledges the cyber-physical character of modern artificial pancreas technologies while remaining grounded in verifiable clinical workflows [35]. Fig. 1 summarises the minimal handling bundle as a compact decision pathway suitable for inpatient use.

## 7. Future Works

Current gaps and their compounding risks emphasise the necessity of a systematic study. Four critical domains: threat modeling voids for patient-integrated systems, evaluation of policy efficacy, longitudinal analysis of DIY ecosystem dynamics, and the case for certified open-source components reveals that the role of regulated open-source components collectively warrants broader and deeper consideration to transform current governance approaches from fragmented and reactive to cohesive and evidence-based. Within each domain, there is a need for refined risk assessment frameworks specific to patient-integrated architecture and account for their unique attack surfaces; comparative analysis to illuminate the evolution of trust mechanisms and adoption patterns; and innovative hybrid options as incentives to reduce the risk of reverse engineering. Addressing these needs in concert would enable proactive governance responses that support DIY AID safety and regulation.

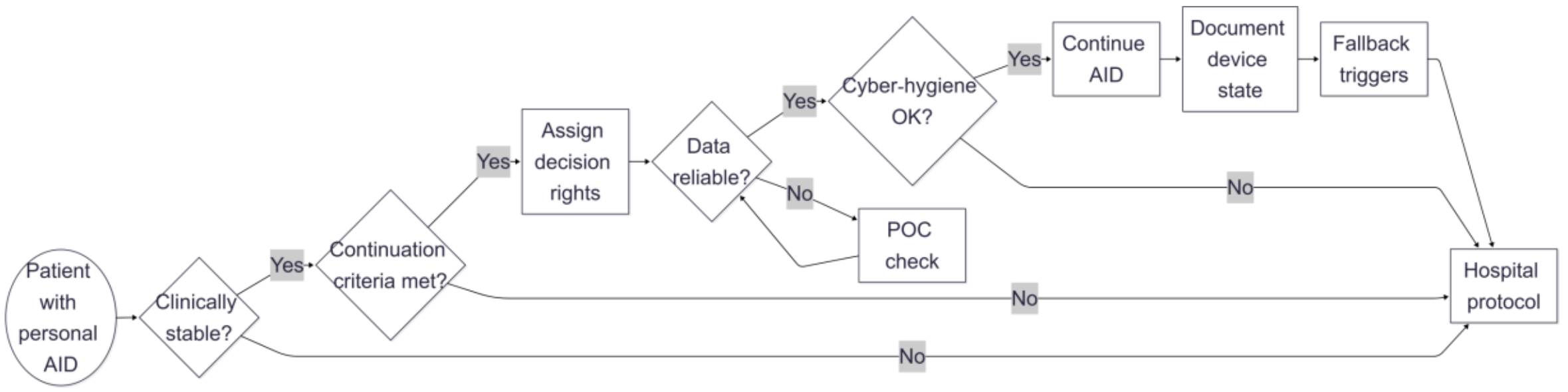

Figure 1. Minimal clinical cyber-safety handling bundle for personal AID systems

## 8. Limitations

This analysis is subject to several important limitations that constrain the generalisability and evidentiary strength of its conclusions. First, much of the empirical literature on DIY artificial pancreas systems is observational, qualitative, or based on self-reports from clinicians and users, introducing selection bias and limiting causal inference regarding safety outcomes [35]. Second, the paper necessarily extrapolates from broader medical device cybersecurity and safety engineering scholarship to characterise cyberbiosecurity risks in artificial pancreatic systems, given the paucity of studies directly examining adversarial or non-adversarial cyber incidents in real-world clinical use. Third, legal and ethical analyses reflect jurisdiction-specific regulatory environments and may not translate uniformly across healthcare systems with different liability regimes or standards of care [34]. Finally, the rapid evolution of both commercial and DIY AID technologies means that specific technical configurations, software practices, and governance norms may change faster than the academic literature can capture, underscoring the need for ongoing reassessment rather than static conclusions. Constraints of this venue format require brevity and expansions of thought are planned for follow-up in future works.

## 9. Conclusion

Artificial pancreas and AID systems illustrate the merging of clinical care, automation, and medical technology, blurring the lines between device safety and cybersecurity. This paper argues that many cyberbiosecurity risks are embedded in current socio-technical workflows, particularly with the rise of DIY artificial pancreas systems, which create a gap between clinical responsibility and technical authority. The paper presents a framework that addresses cyberbiosecurity as a governance and workflow issue, advocating for role clarity, documentation, and interdisciplinary coordination. The proposed minimal clinical cyber-safety handling bundle serves as a practical starting point for harm containment in both regulated and DIY contexts. It highlights the need for systematic research on failure modes and governance mechanisms as AID technologies evolve. Developing evidence-based approaches that balance clinical safety, cybersecurity, and ethical responsibility is crucial for maintaining trust and protecting patients in automated care models. Additionally, while open-source AID systems show promise, they face substantial barriers to wider adoption, despite regulatory progress demonstrating that open-source algorithms can meet safety standards with proper governance. In particular, the EU MDR's requirements for device-specific clinical evaluation, extensive technical documentation, and contractual access to equivalent device files create an evidentiary burden that may dissuade open-source developers from seeking regulatory approval [18]. These constraints underscore that the future of OS-APS depends on both technical capabilities and regulatory environments that can accommodate transparent, modular, community-origin software without compromising safety.